\documentclass{article}

\PassOptionsToPackage{numbers, compress}{natbib}

\pdfoutput=1
\usepackage[preprint]{neurips_data_2022}

\usepackage[utf8]{inputenc} 
\usepackage[T1]{fontenc}    
\usepackage[colorlinks=true]{hyperref}       
\usepackage{url}            
\usepackage{booktabs}       
\usepackage{amsfonts}       
\usepackage{nicefrac}       
\usepackage{microtype}      
\usepackage{xcolor}         
\usepackage{xspace}

\usepackage{enumerate}
\usepackage{cleveref}
\usepackage{subfigure}
\usepackage{colortbl}
\usepackage{bbding}
\usepackage{graphicx}
\usepackage{multirow}
\usepackage{animate}
\usepackage{amssymb}
\usepackage{arydshln}
\usepackage{subfigure}
\usepackage{graphics}
\usepackage{cleveref}
\usepackage{wrapfig,lipsum,booktabs}
\usepackage{etoc}

\crefname{section}{Section}{Sections}
\crefname{theorem}{Theorem}{Theorems}
\crefname{lemma}{Lemma}{Lemmas}
\crefname{equation}{Equation}{Equations}
\crefname{proposition}{Proposition}{Propositions}
\crefname{claim}{Claim}{Claims}
\crefname{appendix}{Appendix}{Appendices}
\crefname{algorithm}{Algorithm}{Algorithms}
\crefname{figure}{Figure}{Figures}
\crefname{table}{Table}{Tables}
\crefname{remark}{Remark}{Remarks}
\crefname{definition}{Definition}{Definitions}
\crefname{equatinon}{Equation}{Equations}
\crefname{corollary}{Corollary}{Corollaries}

\usepackage{diagbox}
\usepackage{etoc}
\etocdepthtag.toc{mtchapter}
\etocsettagdepth{mtchapter}{subsection}
\etocsettagdepth{mtappendix}{none}

\newcommand{\ours}{\texttt{AMOS}\xspace}
\newcommand{\amosct}{\texttt{AMOS-CT}\xspace}
\newcommand{\amosmri}{\texttt{AMOS-MRI}\xspace}

\newcommand{\revise}[1]{\textcolor{black}{#1}}

\title{AMOS: A Large-Scale Abdominal Multi-Organ Benchmark for Versatile Medical Image Segmentation}

\author{Yuanfeng Ji\textsuperscript{\rm 1}, Haotian Bai\textsuperscript{\rm 2}, Jie Yang\textsuperscript{\rm 2}, Chongjian Ge\textsuperscript{\rm 1}, Ye Zhu\textsuperscript{\rm 2}, Ruimao Zhang\textsuperscript{\rm 2*} \\
\textbf{Zhen Li}\textsuperscript{\rm 2*}, \textbf{Lingyan Zhang}\textsuperscript{\rm 3},
\textbf{Wanling Ma}\textsuperscript{\rm 4}, \textbf{Xiang Wan}\textsuperscript{\rm 2*},
\textbf{Ping Luo}\textsuperscript{\rm 1}\thanks{Corresponding authors: \url{pluo@cs.hku.hk}  \& \url{ruimao.zhang@ieee.org} \& \url{lizhen@cuhk.edu.cn }  \& \url{wanxiang@sribd.cn}
},
\\
\textsuperscript{\rm 1} The University of Hong Kong \\
\textsuperscript{\rm 2} Shenzhen Research Institute of Big Data, The Chinese University of Hong Kong (Shenzhen)\\
\textsuperscript{\rm 3} Longgang Central Hospital of Shenzhen, China\\
\textsuperscript{\rm 4} Longgang People Hospital of Shenzhen, China\\}

\begin{document}

\maketitle

\begin{abstract}
Despite the considerable progress in automatic abdominal multi-organ segmentation from CT/MRI scans in recent years, a comprehensive evaluation of the models' capabilities is hampered by the lack of a large-scale benchmark from diverse clinical scenarios.
Constraint by the high cost of collecting and labeling 3D medical data, most of the deep learning models to date are driven by datasets with a limited number of organs of interest or samples, which still limits the power of modern deep models and makes it difficult to provide a fully comprehensive and fair estimate of various methods.
To mitigate the limitations, we present \ours, a large-scale, diverse, clinical dataset for abdominal organ segmentation.
\ours provides 500 CT and 100 MRI scans collected from multi-center, multi-vendor, multi-modality, multi-phase, multi-disease patients, each with voxel-level annotations of 15 abdominal organs, providing challenging examples and test-bed for studying robust segmentation algorithms under diverse targets and scenarios.
We further benchmark several state-of-the-art medical segmentation models to evaluate the status of the existing methods on this new challenging dataset.
We have made our datasets, benchmark servers, and baselines publicly available, and hope to inspire future research. 
Information can be found at \url{https://amos22.grand-challenge.org}.
\end{abstract}
\section{Introduction}
A dense and pixel-precise understanding of abdominal anatomy is of fundamental importance for \revise{computer-aided clinical applications} such as disease diagnosis and radiotherapy planning.
Specifically, accurate abdominal organ segmentation provides crucial information such as the interrelations among the organs as well as individual positions and shapes in the standardized space, which is essential for assisting clinical decision-making. 
With the development of related datasets~\cite{msd,chaos,btcv,densevnet,adb1k}, significant progress has been made in this area.


%
\revise{Nevertheless}, robust segmentation of abdominal organs remains challenging in a real-world clinical setting because of the large variety of organ morphological structures, appearance, and imaging qualities between images acquired from different patients by different scanners.
Some challenges are rooted in the gap between the recent benchmarks and the real-world clinic, which we summarized as follow:
(a) \textit{Small-Scale}: Since the acquisition and annotation of 3D Medical data are incredibly expensive, currently available benchmarks contain only a limited number of data samples or organ annotations, or both.
(b) \textit{Lack of Diversity}: The tremendous cost also restricts previous works to only acquiring data samples from a single-center, phase, scanner, and disease \cite{msd}.
Models trained with homogeneous and inadequate datasets tend to perform catastrophically when the test data distribution shifts under different clinical scenarios. 
To bridge the gap, one more reality-oriented and robust solution is urgently needed to be proposed.

To mitigate the above two challenging limitations, we propose \ours, a comprehensive abdominal organ segmentation dataset with abundant annotations of multi-modality, multi-center, multi-scanner, multi-phase, and multi-disease patients, that covers overall 15 organs. Compared with previous benchmarks, \ours owns its unique contributions to the research community from the following aspects,
(a) Large-Scale: \ours contains 600 Computerized Tomography (CT) / Magnetic Resonance Imaging (MRI) scans with over 74K annotate slices, which is 20$\times$ larger than the previous commonly-used BTCV~\cite{btcv} dataset (3.6K).
(b) Diverse and Clinical: \ours acquires data from the real-world clinical settings, where the patients with different abdominal cancers/abnormalities are tested from eight different CT or MRI scanners at two medical centers. The significant heterogeneity poses higher requirements for algorithm robustness.
(c) Versatile: Despite the multi-organ segmentation, we also show that \ours could potentially serve as a multi-functional dataset for various learning tasks, such as  Out-of-Distribution (OOD) generalization, cross-modality learning, transfer learning, privacy-preserving computation and so on.

With the proposed dataset, we build a new benchmark including existing popular medical segmentation methods. The results prove that current state-of-the-art algorithms fail to make satisfactory performance. Considering the diverse and reality-oriented characteristics owned by \ours, we hope that \ours could serve as a new benchmark to evaluate multi-organ segmentation algorithms in practical applications.
To be more specific, the \textbf{contributions} of our work to the medial segmentation community are as follow:
 \begin{itemize}
     \item We build a new large-scale, diverse, and clinical abdominal organ segmentation dataset of 600 CT/MRI scans, namely \ours, which is comprehensive with 15 organs. To our knowledge, it is the largest dataset of its kind.
     \item We benchmark current baseline methods on this newly built dataset with various evaluation metrics, showing the limitation of existing state of the arts abdominal organ segmentation algorithms.
    \item We carefully design extended experiments to validate that \ours could serve as a versatile dataset for multiple learning tasks.
 \end{itemize}
 

\begin{figure}[t!]
	\centering
	\subfigure
	{
		\animategraphics[width=0.24\linewidth,height=0.2\linewidth, autopause, poster=1]{2}{sources/figures/animates/amos_159/image_}{00011}{000015}\hspace{-1.5mm}
	}
    \vspace{-1mm}
	\subfigure
	{
		\animategraphics[width=0.24\linewidth,height=0.2\linewidth, autopause, poster=1]{2}{sources/figures/animates/amos_405/image_}{00007}{00011}\hspace{-1.5mm}
	}
  	\vspace{-1mm}
	\subfigure
	{
		\animategraphics[width=0.24\linewidth,height=0.2\linewidth, autopause, poster=1]{2}{sources/figures/animates/amos_301/image_}{00008}{00011}\hspace{-1.5mm}
	}
  	\vspace{-1mm}
	\subfigure
	{
		\animategraphics[width=0.24\linewidth,height=0.2\linewidth, autopause, poster=1]{2}{sources/figures/animates/amos_446/image_}{00007}{00010}
	}
  	\vspace{-1mm}
	\subfigure
	{
		\animategraphics[width=0.24\linewidth,height=0.2\linewidth, autopause, poster=1]{2}{sources/figures/animates/amos_600/image_}{00014}{00017}\hspace{-1.5mm}
	}
	\subfigure
	{
		\animategraphics[width=0.24\linewidth,height=0.2\linewidth, autopause, poster=1]{2}{sources/figures/animates/amos_558/image_}{00014}{00018}\hspace{-1.5mm}
	}
	\subfigure
	{
		\animategraphics[width=0.24\linewidth,height=0.2\linewidth, autopause, poster=1]{2}{sources/figures/animates/amos_540/image_}{00014}{00018}\hspace{-1.5mm}
	}
	\subfigure
	{
		\animategraphics[width=0.24\linewidth,height=0.2\linewidth, autopause, poster=1]{2}{sources/figures/animates/amos_535/image_}{00006}{00010}
	}
	\caption{Example annotated slices from \ours dataset. Watch the animations by \textcolor{red}{\textbf{clicking}} them (Not all PDF readers support playing animations. Best viewed with Acrobat/Foxit Reader). The top and bottom two rows show the CT and MRI slices acquired from different scanners, respectively.}
	\label{fig:intro:anime}
\end{figure}

\section{Related Work}

\setlength{\tabcolsep}{8pt}
\begin{table}[t!]
	\scriptsize
	\centering
	\begin{tabular}{lccccccccc} \hline
		Abdomen Organ Dataset & \#Organs &\#Scans &\#Slices & \# Anns per Scan &Modality & Region & Year   \\ \hline
		MSD-Liver$^{\dagger}$ \cite{msd} & 1 & 201 &29,402 & 3410K & CT & France & 2018 \\
		MSD-Spleen$^{\dagger}$ \cite{msd} & 1 & 61 &1,563 &40K & CT & United States &2018\\
		MSD-Prostate$^{\dagger}$ \cite{msd} & 1 & 48 & 712 & 15K & MRI & Netherlands &2018 \\
		MSD-Pancreas$^{\dagger}$ \cite{msd} & 1 & 420 & 13,141 & 144K  & CT & United States &2018\\
		Kits$^{\dagger}$ \cite{kits} &1 &300 &23,337 &997K & CT & United States &2019  \\ \hline 
		BTCV$^{\dagger}$ \cite{btcv} & 13 & 50 &3,629 &431K & CT & United States & 2015   \\
		Chaos$^{\dagger}$ \cite{chaos} & 4 & 80 &1,989 &52K  & CT\&MRI & Turkey & 2019 \\
		DenseVNet \cite{densevnet} & 8 & 90 &9,246 &1430K & CT & United States & 2018  \\
		AbdomentCT-1K$^{\dagger}$ \cite{adb1k} & 4 & 1,112 &34,497 &3412K & CT &Worldwide & 2021   \\
		Word \cite{word} & 16 & 150 & - & - & CT & China & 2021   \\  \hline
		\ours (ours) & 15 & 600 &74,026 &9952K  & CT\&MRI & China & 2022 \\  
 	\hline
	\end{tabular}
	\vspace{0.1cm}
	\caption{Comparison of \ours dataset with other conventional abdominal organ segmentation datasets. $^{\dagger}$ indicates that the value of the dataset is the estimate from released partial data. - means unavailable estimation due to data inaccessibility.}
	\label{tab:amos:comparsion}
\end{table}

\paragraph{Abdominal organ segmentation datasets}
As summarized in \cref{tab:amos:comparsion}, existing abdominal organs segmentation datasets vary in size, modality as well as the number of annotations.
For example, for the single-organ segmentation dataset, the MSD~\cite{msd}, considered as the most commonly-used one, provides annotations of the single organ individually (e.g., the liver, spleen, prostat, and pancreas), with data \revise{sizes} ranging from tens to hundreds.
For the multi-organ segmentation dataset, BTCV~\cite{btcv} takes the pioneering step to provide 50 CT scans covering 13 types of organ annotations.
Besides, DenseVNet~\cite{densevnet} improves the BTCV dataset via involving additional data, forming a total dataset consisting of 90 scans with eight organ annotations.
Compared to the single-modality dataset, the Chaos Dataset~\cite{chaos} provides multi-modality information, including both 20 CT scans for the liver and 20 MRI data with four organs.
More recent works aim to include more samples in the specific dataset, e.g., AbdomenCT-1K~\cite{adb1k} provides 1,112 scans with four organs, while Word~\cite{word} contains 150 CT samples to cover 16 kinds of abdominal organs. 
Unlike the above works, our \ours contains 600 CT/MRI scans with 15 types of organ annotation, making it the most comprehensive and \revise{diverse} benchmark of its kind to date.
Besides, considering the underrepresentation problem in clinical datasets, our proposed \ours from the Asian sub-population will be a valuable resource to the existing pool.
We refer the readers to \cref{app:datasets} for more detailed information on the mentioned datasets.

\paragraph{Methods for single abdominal organ segmentation}
The single organ segmentation has been the dominant task for decades, where numerous solutions have been developed~\cite{hdenseunet,nnunet,coarse2fine,cascade_unet,focusnet,uxnet,c2fnas}.
For example, to precisely segment the liver and tumor, H-DenseUNet~\cite{hdenseunet} designs a hybrid 2D/3D network for better features extraction.
By proposing a self-configured framework based on the naive UNet~\cite{unet}, nnUNet~\cite{nnunet} achieves superior performance on segmenting the liver, spleen, kidney, as well as pancreas,
\revise{the approach can be easily adapted to multi-organ segmentation tasks}.
Besides, to resolve the challenges of small target organs, a series of works~\cite{coarse2fine,cascade_unet,focusnet} adopt the cascaded structures, where the networks are designed in a coarse-to-fine manner.
More recent works~\cite{uxnet,c2fnas} replace the manual designs by the neural network searching for optimal segmentation architectures to achieve better performance with fewer parameters.

\paragraph{Methods for Abdominal multi-organ segmentation}
For multi-organ segmentation, where multiple organs are segmented simultaneously, networks should be designed to own more powerful ability of discriminating the pixel-wise features.
OAN \cite{oan} designs a fusion network that takes 2D multi-view images as input and reconstructs the 3D segmentation result finally.
DenseVNet~\cite{densevnet} proposes a dense 3D network for performance improvement.
Besides, due to the data insufficiency, a \revise{series} of works~\cite{psnn,pipo,dodnet} propose different training paradigms and achieve multi-organs segmentation using partially label annotations from single organ datasets.
In this group of work~\cite{nnformer,swinunetr,unter,cotr}, the features patch or image patch are treated as tokens, which are used to conduct efficient non-local context modeling among arbitrary positions, making them achieve SOTA performance on the most popular dataset~\cite{btcv}.
However, we empirically found that developing and validating methods with limited data still hinders the potential power of modern deep models, leading to unfair/inaccurate comparisons and estimates of different methods.

\section{AMOS} \label{sec:amos}

\ours collects both the CT and MRI image data from anonymous patients in clinical medical centers, with abundant segmentation annotations. In this section, we will detail the data curation process for both the images and annotations, as well as provide recommendations for potential usages. Additional details, including acquisition, annotation and so on, can be found in \cref{app:amos}.

\subsection{Dataset Construction} \label{sec:data_construction}
\paragraph{Data Overview}
Following the standard clinical acquisition protocols, the CT/MRI data are collected from 600 patients, who are diagnosed with abdominal tumors/abnormalities at Longgang District People's Hospital and Longgang District People's Hospital, via one of the eight machines as shown in Table~\ref{tab:amos:data_collection}.
Both scanner-generated DICOM and diagnosis reports are collected, de-identified, and stored securely. All data contributions to this study have been approved by the Research Ethics Committees of Longgang District People's Hospital (reference number: 2021077) and Longgang District Central Hospital (reference number: 2021ECJ012).

\begin{figure*}[t!]
	\centering
	\includegraphics[width=\linewidth]{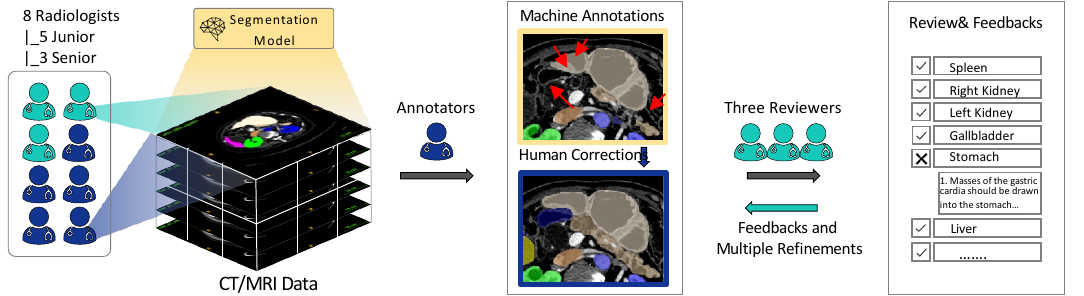}
	\caption{Annotation workflow of AMOS. The coarse annotations automatically labeled by pre-trained segmentors will be further refined by human annotators for multiple times, including 5 junior radiologists for the initial stage and 3 senior specialists for the second checking stage.}
	\label{fig:amos:annotation_workflow}
\end{figure*}

\paragraph{Data Collection} \ours is designed to facilitate abdominal multi-organ segmentation in a more diverse, clinical, and complex scenarios. To meet this purpose, we cautiously select data, generated by two institutes from 2018 to 2021, based on the following criteria:
1) Patients should be diagnosed with abdominal tumors/abnormalities, while the ones with normal abdomen will be excluded.
2) The imaging quality of the scanned data should be high-quality enough for the radiologists to review and annotate.
3) To ensure the data diversity, the collected samples should be derived from different scanners, as well as different scanning stages.
4) The scan data should cover as many of the specified abdominal organs as possible.
To this end, we collected a total of 500 CT and 100 MRI scans from 600 unique patients, covering 15 organ categories, including spleen, right kidney, left kidney, gallbladder, esophagus, liver, stomach, aorta, inferior vena cava, pancreas, right adrenal gland, left adrenal gland, duodenum, bladder, prostate/uterus.

\paragraph{Data Annotation} Considering the expensive cost of labeling 3D data, we follow~\cite{adb1k,step} to propose a semi-automatic annotation workflow as shown in Figure~\ref{fig:amos:annotation_workflow}. There are two stages for annotation, including the coarse labeling stage and the refinement stage.
Specifically, a few samples (50 CT and 20 MRI scans) are initially annotated with human labors. Then, we conduct the training process with the annotated scans on several representative models, e.g., 3D-UNet~\cite{unet} and VNet~\cite{vnet}. The pre-trained segmentors are utilized to pre-label the remaining scans automatically and coarsely. 
To this end, we finish the coarse labeling stage.
For the second stage, five well-trained junior radiologists are employed to check and revisit the segmentation results on a case-by-case basis. To further reduce errors/biases, three senior radiologists with more than 10 years of clinical experience are responsible for the final validation. Typically, they will conduct the annotation review, errors revision, and feedback distribution to improve the annotation quality. The overall process will be iterated several times in the second stage for the final consensus on the well-labeled annotations. Such interactive checking reduces the possible bias caused by individual annotators.
We illustrate our semi-automatic annotation workflow in \cref{fig:amos:annotation_workflow}. It is evident that human refinements and corrections are required to achieve high-quality annotations.
Overall, \ours annotates 500 CT and 100 MRI scans, resulting into two sub-datasets, i.e., \amosct and \amosmri.
We refer the readers for more technical details in the \cref{app:amos:annotate}.

\setlength{\tabcolsep}{15pt}
\begin{table}[t!]
\scriptsize
\centering
\begin{tabular}{llcccccc}
\hline  & Scanner & Domain & \#Train & \#Validate & \#Test &  \# Total  \\ \hline 
\amosct & Aquilion ONE & A & 82 & 27 & 27 & 136 \\
$-$ & Brilliance16 & B  & 68 & 23 & 23 & 114 \\
$-$ & Somatom Force & C & 50 & 50 & 28 & 128 \\
$-$ & Optima CT660 & D  & $-$ & $-$ & \cellcolor{gray!15}64 & 64\\
$-$ & Optima CT540 & E  &$-$ & $-$ & {\cellcolor{gray!15}}58& 58 \\ \cdashline{4-7}[0.5pt/5pt]
& & &200 &100 &200 &500 \\
\hline
\amosmri & Ingenia & F & $-$ & $-$ &\cellcolor{gray!25} 29 & 29 \\
$-$ & Prisma & G & 33 & 20 &11 & 64 \\
$-$ & Signa HDe & H & 7 & $-$ & $-$ &  7 \\ \cdashline{4-7}[0.5pt/5pt]
& & &40 &20 &40 &100 \\
\hline
\end{tabular}
\vspace{0.1cm}
\caption{Partition of the \ours dataset. In addition to the conventional train-to-training in-distribution setting, we extend the evaluation setting to out-of-distribution data, where performance is also measured on unseen test data (marked as gray cell)}.
\label{tab:amos:data_collection}
\end{table}

\paragraph{Data Splits}
%
%
%
Distribution shift, where the training distribution differs from the test distribution, is ubiquitous in medical applications.
In practice, it is often the case that the testing data (a.k.a., target domain) can  \revise{differ from} training data (a.k.a., source domain) in a variety of ways, such as imaging protocol, device vendors, and patient populations.
Such a domain shift always significantly degrade the performance of the developed model.
In \ours, we consider the domain shift caused by the device vendors, where domains are the image acquisition scanner. Besides, the model needs to generalize to data from a new scanner that is disjoint with the training set.
Specifically, as shown in \cref{tab:amos:data_collection}, we make data split according to their performed scanner.
For the \amosct, 378 unique abdominal scans from three scanners are randomly split into 200 (n=82, domain A; n=68, domain B; n=50, domain C), 100 (n=27, domain A; n=23, domain B; n=50, domain C), 78 scans (n=27, domain A; n=23, domain B; n=28 domain C) for training, validation, testing (In Distribution, ID) respectively, the other 122 cases (n=64, domain D; n=58, domain E) from other two scanners serve as unseen test data and consist to the testing (Out of Distribution, OOD) split.
The same split strategy is adopted to \amosmri, resulting in 40 scans for training, 20 scans for validation and 29 scans for testing (OOD), and 11 scans for testing (ID).

\paragraph{Data Distribution} All the scanned samples and segmentation annotations are distributed in multiple formats (e.g., DICOM, NIFTI) to ensure their applicability in both the research community and clinical scenarios. Patient-protected health information (PHI) metadata was removed from DICOM files. For additional details on distribution and maintenance, please refer to \cref{app:amos:dist}.

\subsection{Dataset Statistics}

\begin{figure}[!h]
\vspace{-0.2cm}
	\centering
	\subfigure[Top-ten most frequent diseases and diseased organs.]
	{ 
		\includegraphics[width=0.475\linewidth,height=0.25\linewidth]{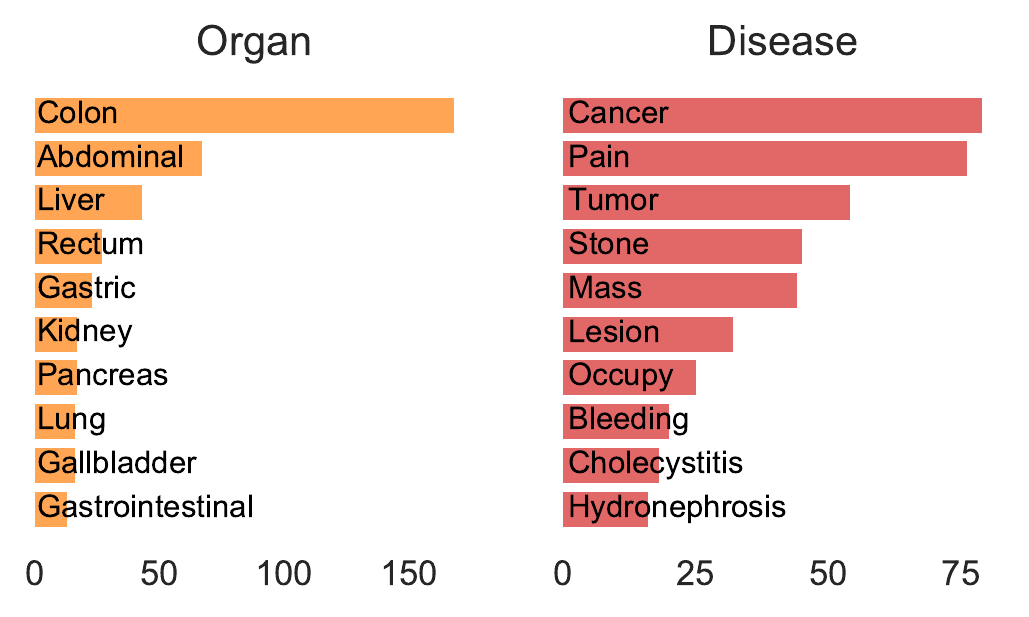}
	}
	\subfigure[Number of annotated voxels per category]
	{
		\includegraphics[width=0.475\linewidth,height=0.25\linewidth]{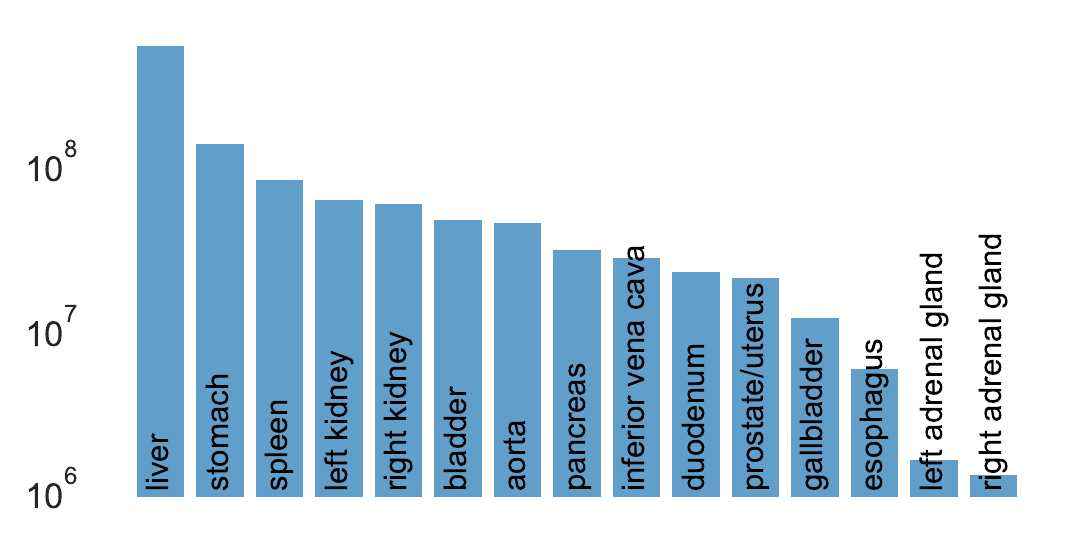}
	}
    \caption{Statistics on data targets as well as data annotation, reflecting that AMOS is a clinical, highly diverse data set. The x-axis units of both figures are counts
    }
	\label{fig:amos:statistics}
	\vspace{-0.5cm}
\end{figure}

\paragraph{Cohort Statistics} 
To analyze the cohort characteristics of \ours, we summarize the diagnoses of the collected patients from two aspects (i.e., affected organ and disease type) based on the patient reports. The word frequency results, in ~\cref{fig:amos:statistics}(a), show that \ours spreads over a large range of diseases and organs, validating the diversity and variability of our dataset. Due to the page limit, more cohort characteristics (e.g., sexual and age information) are available in \cref{app:amos:stat}.


\paragraph{Annotation Statistics} We also conduct the statistical measurements to analyze the semantic label distribution in \ours. As shown in~\cref{fig:amos:statistics}(b), the annotation of \ours naturally has a long-tail distribution. For example, the scale of liver annotation is about $\mathrm{200}\times$ larger than the adrenal gland. This natural/reality-oriented long-tail distribution property makes \ours more challenging for precise multi-organ segmentation.



\setlength{\tabcolsep}{9pt}
\begin{table}[!t]
\centering
\scriptsize
\begin{tabular}{l|cc|cccc} \hline
\diagbox[height=9pt]{Target}{Source}& BTCV~\cite{btcv}&\ours &MSD-Spleen~\cite{msd} &MSD-Liver~\cite{msd} &MSD-Pancreas~\cite{msd} & Kits~\cite{kits} \\\hline
BTCV &83.62 &\textbf{84.47} & \_ & \_ & \_ & \_\\
\ours &79.74 &\textbf{90.83} & \_ & \_ & \_ & \_\\
MSD-Spleen &91.10 & 94.73 &\textbf{95.39} & \_ & \_ & \_ \\
MSD-Liver &92.32 &95.40 &\_ &\textbf{96.85} & \_ & \_ \\
MSD-Pancreas &66.32 &71.76 &\_ &\_ &\textbf{86.68} & \_ \\
Kits &80.21 &88.33 &\_ &\_ &\_ &\textbf{97.47} \\
\hline
\end{tabular}
\vspace{0.1cm}
\caption{The model pretrained on the \ours dataset shows superior transfer performance compared to those on the BTCV and MSD dataset, which implies our superior data capacity and quality.}
\label{tab:amos:quality}
\vspace{-0.5cm}
\end{table}

\paragraph{Dataset Comparison} 
In this part, we compare the data statistics of our dataset to those of commonly used abdominal organ segmentation datasets. Since not all annotations of these datasets are available to the public, some statistics marked with $^{\dagger}$are estimated using a subset of the dataset.
The overall comparisons are conducted from the aspects of data scale and diversity.
First, we compare the data scale between \ours and other representative segmentation datasets in~\cref{tab:amos:comparsion}.
Specifically, \ours contains 74K annotated slices, which is $\mathrm{2.2}\times$ larger than the second-largest dataset AdbomentCT-1K, and $\mathrm{20}\times$ larger than BTCV. For the number of scanned samples, our \ours is significantly larger than most datasets except for AdbomentCT-1K. Nevertheless, \ours still takes advantage of organ numbers, annotation numbers in a single scan, and modality numbers over AdbomentCT-1K.
Second, for annotation diversity comparison, \cref{tab:spacing} in the \cref{app:amos:stat} shows that the min/max/median values of slice spacing and size range are [0.8/6.0/5.0] and [40/553/115], respectively, which spreads diversely than other abdominal multi-organ datasets.
Besides, in term of the organs volume, we visualize the distributions of Chaos, BTCV, AdbomentCT-1K and \ours in \cref{fig:amos:comparsion:volume} in the Appendix. Results reveal that \ours's distribution almost covers the others, validating its high diversity. Such a property puts a higher requirement on the model's capacity and perception abilities.

To further validate the data diversity and evaluate the annotation quality, we propose an experimental comparison. Specifically, we first pre-train two UNets based on \ours and its \revise{closest dataset}, BTCV, respectively. The pre-trained models will be employed to infer on multiple target datasets, e.g., MSD-Spleen, MSD-Liver, MSD-Pancreas, and Kits. Given two datasets for comparison, the performance are evaluated on 
their mutual classes. As shown in \cref{tab:amos:quality}, The model trained on \ours dataset attains 94.73\%, 95.40\%, 71.76\%, 88.33\% mDice scores on the target dataset MSD-Spleen, MSD-Liver, MSD-pancreas, Kits dataset, respectively, which significant outperforms the one trained on the BTCV dataset (91.10\%, 92.32\%, 66.32\%, 80.21\%), being slightly worse than MSD-pretrained models (95.39\%, 96.85\%, 86.68\%, 97.47\%).
Besides, the model trained on \ours achieves 84.47\% mDice on the BTCV validation set, surpassing the model trained on BTCV training set, which indicates the annotation quality and data diversity of \ours facilitate the models in learning better representations.

\subsection{Evaluation Metrics} 
\vspace{-0.2cm}
Following the previous work~\cite{btcv}, we leverage two classical medical segmentation metrics for evaluation, i.e.,  dice similarity coefficient (DSC), and normalized surface dice (NSD)~\cite{nsd}. DSC score is a popular segmentation metric applied to a variety of segmentation tasks, while the NSD score can provide supplementary information on segmentation quality for evaluating the precision of segmentation boundaries. Specifically, we first calculate the category-wise performances based on the above two metrics and then average the values to obtain the overall evaluation \revise{(i.e., mDice and mNSD)}.
For both metrics, a \revise{higher} score indicates a better result. Besides, considering computational efficiency, we encourage future works to provide information on the model capacity (the number of parameters) and computational cost (GFlops).


\vspace{-0.2cm}
\subsection{Potential Usages} 
\vspace{-0.2cm}
Despite the multi-organ segmentation, \ours provides multi-fold and abundant information, enabling other potential usages in the research community. For instance, in~\cref{tab:amos:data_collection}, we have shown that our \amosct and \amosmri consist of five and three domains, respectively. It motivates us to conduct corresponding extensions, e.g., OOD generalization, cross-modality learning and transfer learning. 
Besides, under the general framework of domain shifts, \ours can also be easily adapted to domain adaption and sub-population shift problems. Since each domain in \ours can be treated as a separate client, it is also suitable for developing Privacy-Preserving Computation techniques, such as Federated Learning, which are crucial in medical applications.
We welcome community contributions towards exploring more potential usages of \ours.

\vspace{-0.2cm}
\section{Experiments} \label{sec:experiments}
\vspace{-0.2cm}
\setlength{\tabcolsep}{3pt}
\begin{table}[!t]\centering
	\scriptsize
	\begin{tabular}{lcc|cccc|cccc} \hline
		\multirow{2}{*}{Model} &\multirow{2}{*}{Params(M)} &\multirow{2}{*}{Flops(G)} &\multicolumn{2}{c}{CT-Val} &\multicolumn{2}{c|}{CT-Test} &\multicolumn{2}{c}{MRI-Val} &\multicolumn{2}{c}{MRI-Test} \\
		& & &mDice(\%)&mNSD(\%) &mDice(\%) &mNSD(\%) &mDice(\%) &mNSD(\%) &mDice(\%) &mNSD(\%) \\\hline
		UNet \cite{unet} &31.18 &680.31 &\textbf{88.87} &\textbf{79.87} &\textbf{89.04} & \textbf{78.32} &\textbf{85.59} &\textbf{80.56} &\textbf{67.63} &\textbf{59.02} \\
		VNet \cite{vnet} & 45.65&849.96&81.96 &67.94 &82.92 &67.56 &83.86 &78.0 &65.64 &57.37 \\ \hline
		CoTr \cite{cotr} & 41.87&668.15& 77.13 & 64.15 &80.86 &66.31 &77.5 & 70.1&60.49 &51.18 \\
		nnFormer \cite{nnformer} & 150.14&425.78&85.63 &74.15 &85.61 &72.48 &80.6 &74 &62.92 &54.06 \\ \hline
		UNetr \cite{unter} & 93.02 &177.51&78.33 &61.49 &79.43 &60.84 &75.3 &65.3 &57.91 &47.25 \\
		Swin-UNetr \cite{swinunetr}  & 62.83 &668.15&86.37 &75.32 &86.32 &73.83 &75.7 &65.8 &57.5 &47.04 \\ \hline
	\end{tabular}
	\vspace{0.1cm}
	\caption{Overall results of six baselines method on the \amosct and \amosmri datasets, respectively. The FLOPS/the number of parameters is estimated by using the [1 × 128 × 128 × 128] \revise{patch as model input}. The class-wise scores can be found in \cref{app:exp:res}.}
	\label{tab:exp:baseline}
	\vspace{-2em}
\end{table}
%

In this section, we perform experimental validation on various datasets to investigate the superior properties of our proposed \ours. First, we introduce the experimental settings, including the selected baseline methods and implementation details. Then, experiments are carefully designed to report the performance from multiple perspectives, with various baseline methods and datasets.

\subsection{Experimental Settings} \label{sec:experiments_settings}

\paragraph{Baseline methods} We select recent state-of-the-art segmentors as baseline methods, including the CNN-based methods (e.g., UNet~\cite{unet}), Transformer-based methods (e.g., UNetr~\cite{unter}, Swin-UNetr~\cite{swinunetr}), and the hybrid ones (e.g., Cotr~\cite{cotr}, nnFormer~\cite{nnformer}).
%
%
Unless specified otherwise, we follow the default training and testing configurations provided in the published papers or the released codebases to calculate the performances.

%
\paragraph{Implementation details} In this work, we use the Pytorch toolkit~\cite{pytorch} to conduct all experiments on one single NVIDIA V100 GPU. The nnUNet~\cite{nnunet} codebase is adopted for benchmark implementation. For more training and testing details, we refer the readers to \cref{app:exp:imp}.

\subsection{Benchmark results} 
\vspace{-0.2cm}
To comprehensively evaluate different segmentors on \ours, we first train representative models on the training split, and then report the corresponding performances in~\cref{tab:exp:baseline}. The computational costs, i.e., floating-point operations per second (Flops) and the number of parameters, are also reported for efficiency comparisons. 
Surprisingly, we find that UNet consistently outperforms other recently developed methods for a certain margin, achieving 88.87\% and 85.59\% mDice, and 79.87\% and 80.56\% mNSD on \amosct and \amosmri, respectively. 
Besides, it's found that the Transformer-based models take no obvious performance advantages over the CNN-based models, with overhead computational consumption. For example, UNet with much less parameters surpasses nnFormer, Swin-UNetr over $2\sim3$\% mDice and $4\sim5$\% mNSD scores on \amosct.
To further analyze the segmentation performance, we show the detailed category-wise scores in \cref{app:exp:res}. Results show that both the Transformer-based methods and CNN-based methods are able to perform well on the large organs (e.g., spleen, liver, and kidney), but still poorly on small organs (e.g., duodenum and adrenal gland). It indicates that the more fine-grained features needed to be captured to assist the segmentors in pixel-wise understanding and precise segmentation.
Finally, we have also observed a significant gap between mDice and mNSD as previously mentioned by~\cite{adb1k}. Future segmentation algorithms should move toward improving the boundary segmentation accuracy as well as the continuity to achieve better segmentation results.

\subsection{Generalization Results}
\vspace{-0.2cm}
\label{sec:experiments:general}

To analyze the generalization capabilities of different segmentors, we evaluate their performances on the testing (ID) and testing (OOD) splits, respectively. As shown in~\cref{tab:exp:baseline:ood}, the performances between two splits vary significantly on \amosmri, but almost keep consistent on \amosct. For example, for \amosmri, UNet achieves an 86.05\% mDice on the testing (ID) set, but only 64.07\%($-\mathrm{21.25}\%$) mDice on the testing (OOD) set. Similarly, the corresponding performance gap for nnFormer is $-\mathrm{17.68}\%$ mDice (80.6\% \textit{v.s.} 62.92\%). The results demonstrate that there is indeed a data distribution shift caused by collecting MRI samples from different scanners that hinders the model from performing consistently on two sub-sets. However, for \amosct, the performance gap between testing ID and OOD set disappears (e.g., 88.87\% \textit{v.s.} 89.04\% mDice for UNet). It is because the CT samples collected by different scanners in our work are almost indistinguishable from each other since the scans follow the same imaging standards, including the imaging protocols, intensity, and so on. The detailed properties of the collected data, shown in \cref{app:amos:stat}, also validate the imaging intensity is consistent in \amosct but distinguishable in \amosmri.
Due to the different data distributions in \amosmri subsets, we encourage the research community to explore \amosmri for more OOD generalization tasks.

\subsection{Extended explorations on \ours} Since \ours \revise{contains} more modalities and abundant annotations than other works, we are able to additionally investigate the following properties of \ours, which haven't been explored by previous abdominal segmentation datasets.

\paragraph{Cross-Modality learning}
\begin{wraptable}{r}{6cm}
\vspace{-0.3cm}
\centering
\scriptsize
\begin{tabular}{lcc} \hline
Training data &CT-Val &MRI-Val \\ \hline
Individual (200CT / 40MRI) & 88.87/79.97 & 85.59/80.56 \\
Joint (200CT+40MRI) &89.42/80.78 & 87.73/82.72 \\ \cdashline{1-3}[0.5pt/5pt]
Joint (160CT+40MRI) &89.12/80.19 & --/-- \\
Joint (10CT+30MRI) & --/-- & 86.28/80.96 \\ \hline
\end{tabular}
\caption{\footnotesize Cross modality learning results. The performance is reported in mDice/mNSD.}
\label{tab:exp:corssmodality}
\vspace{-0.4cm}
\end{wraptable}
%
Though CT and MRI contain internal body information, the information that they are good at capturing is different. For instance, CT shows more potential in imaging organs and skeletal structures, while MRI especially focuses on more fine-grained organ details and soft tissues. 
An intuitive motivation is that the data from two modalities (i.e., CT and MRI) could complement each other for better clinical diagnosis.
Here, we conduct experiments to improve the segmentation performance via cross-modality learning. Specifically, we first train the UNet model on three datasets (\amosct, \amosmri, and their joint), and then evaluate its performance on each modality. Results in~\cref{tab:exp:corssmodality} show that the jointly trained model consistently improves the individually trained model on \amosct (i.e.,+0.55\% mDice and +0.81\% mNSD) and \amosmri (i.e., +2.14\% mDice and +2.16\% mNSD).
To mitigate the effect that the improvement is caused by more training data, we conduct two comparative experiments by selecting training samples randomly (i.e., 160CT+40MRI, 10CT+30MRI). The improvement could also be observed, which validates the effectiveness of cross-modalities training.


\setlength{\tabcolsep}{4pt}
\vspace{-0.4cm}
\begin{wraptable}{r}{8cm}
\centering
\scriptsize
\begin{tabular}{lcccc}\hline
&\multicolumn{2}{c}{Train from scratch} &\multicolumn{2}{c}{Pretrained with \ours} \\ \cline{2-5}
&mDice(\%) &mNSD(\%) &mDice(\%) &mNSD(\%) \\ \hline
MSD-Liver~\cite{msd} &80.31 &57.40 &81.97$\uparrow$ &59.14$\uparrow$ \\
MSD-Spleen~\cite{msd} &95.39 &88.40 &96.40$\uparrow$ & 90.12$\uparrow$ \\
MSD-Pancreas~\cite{msd} &66.71 &47.89 &66.17 &48.32 \\
MSD-Prostate~\cite{msd} &80.64 &56.75 &81.47$\uparrow$ &58.73$\uparrow$ \\
Kits~\cite{kits} &88.52 &81.97 &89.93$\uparrow$ &83.62$\uparrow$\\
SegTHOR &89.66 &74.77 &90.91$\uparrow$ &75.29$\uparrow$ \\ \hline
MSD-Cardiac~\cite{msd} &93.89 &86.58 &93.67 &85.89 \\
MSD-HepaticVesse~\cite{msd} &70.21 &61.5 &70.36 &61.79 \\
ACDC~\cite{acdc}&92.30 &93.64 &92.45 &94.05 \\
Covid19-Seg~\cite{covidseg} &72.53 &56.22 &72.68 &56.51 \\
\hline
\end{tabular}
\caption{\footnotesize  Transfer learning results on 10 target datasets.}
\label{tab:exp:transfer}
\vspace{-0.5cm}
\end{wraptable}


\vspace{0.2cm}
\paragraph{Transfer Learning} 
Models trained on the large-scale and diverse dataset are usually supposed to have stronger transfer learning abilities to perform well on the related sub-sets. We conduct experiments in this part to validate how \ours benefits models in transfer learning scenarios.
Taking UNet as the baseline method, we train the models with two training configurations: 1) the model is trained on the specific dataset from scratch, 2) the model is first pre-trained on \ours and then fine-tuned on the other datasets. The performances are evaluated on 10 segmentation tasks, including six related datasets that contain organs in \ours, and four unrelated ones. The overall tasks are suitable for evaluating the transfer learning abilities of \ours, since they contain both the seen and unseen domains.
We detail these datasets and the corresponding experimental details in the \cref{app:exp:res}.
Results shown in~\cref{tab:exp:transfer} demonstrate that model trained on \ours consistently outperforms the one in train-from-scratch manner on 5/10 benchmarks, while performing comparably on the other 5/10 tasks. It indicates when the sub-sets are related to \ours, the representations trained by \ours are transferable and beneficial. On the other hand, when the sub-sets domains are significantly different from \ours, e.g., MSD-Cardiac, \ours's representations will not cause the deleterious effects.

\setlength{\tabcolsep}{6pt}
\renewcommand{\arraystretch}{1}
\begin{table}[!tp]\centering
	\scriptsize
	\begin{tabular}{l|cccc|cccc} \hline
		\multirow{2}{*}{Model} &\multicolumn{2}{c}{CT-Test (ID)} &\multicolumn{2}{c|}{CT-Test (OOD)} &\multicolumn{2}{c}{MRI-Test (ID)} &\multicolumn{2}{c}{MRI-Test (OOD)} \\
		&mDice(\%)&mNSD(\%) &mDice(\%) &mNSD (\%) &mDice (\%) &mNSD (\%) &mDice (\%) &mNSD (\%) \\\hline
		UNet \cite{unet}  &\textbf{88.04} &\textbf{79.92}& \textbf{89.67} & \textbf{78.11} &\textbf{86.05} &\textbf{82.79} &\textbf{64.70} &\textbf{54.09} \\
		VNet \cite{vnet}  &81.60 &69.30 &83.76 &66.45 &85.06 &80.93 &62.57 &52.53 \\ \hline
		CoTr \cite{cotr}  &79.32 & 67.54 &81.83 &65.53 &78.62 & 71.94&57.44 &46.86 \\
		nnFormer \cite{nnformer}  &84.43 &74.03 &86.35 &71.50 &80.62 & 74.86 &60.29 & 49.96 \\ \hline
		UNetr \cite{unter}  &77.86 &62.71 &80.43 &59.65 &74.36 &65.33 &55.44 &43.69 \\
		Swin-UNetr \cite{swinunetr} &84.91 &74.96 &87.20 &73.10 &75.89 &67.77 &54.35 &42.59 \\ \hline
	\end{tabular}
	\vspace{0.1cm}
	\caption{The in-distribution (ID) v.s. out-of-distribution (OOD) performance of models trained with empirical risk minimization. The OOD test set are drawn from data distinct from training data as described in \cref{sec:data_construction}, while the ID comparison are draw from the training distribution. \amosmri dataset show significant performance drop due to to distribution shift , with substantially better ID performance than OOD performance. While the \amosct do not show significant performance gap difference between the ID and OOD test set. We analysis the results in \cref{sec:experiments:general}.
	}
	\label{tab:exp:baseline:ood}
	\vspace{-0.6cm}
\end{table}

\section{Discussion  \& Conclusion} \label{sec:conclusion}
In this work, we present a large-scale, diverse, clinical dataset for abdominal multi-organ segmentation, termed \ours.
Thanks to its abundant information, we are able to explicitly focus on measuring algorithm performance fairly and comprehensively.
The extensive experiments on \ours show that our dataset can not only serve as a general multi-organ segmentation benchmark, but also assist the research community in extended applications, including OOD generalization, cross-modality learning, and transfer learning, which haven't been explored by previous works. 
We believe that our work provides an important step toward the dense, pixel-precise abdominal anatomy understanding and other future explorations.
%
For the limitation, first, all data were acquired from patients who received an abdominal scan at two hospitals, which is a subgroup of the Asian population. Second, although data were collected from different scanners, both scanners were from the two hospitals, which may affect model performance on data acquired from the different medical centers. 
Additionally, there are other anatomical structures in the scans (e.g., lung, heart), which is not densely annotated but can help medical workflow.
In future work, we will look to curate data from a more wide patient pool and multi-center scanners and add annotations for additional organs.
Besides, we carefully solve privacy and social effects by strictly obtaining the Research Ethics Committees Approvement.



\begin{thebibliography}{10}

\bibitem{msd}
Michela Antonelli, Annika Reinke, Spyridon Bakas, Keyvan Farahani, Bennett~A
  Landman, Geert Litjens, Bjoern Menze, Olaf Ronneberger, Ronald~M Summers,
  Bram van Ginneken, et~al.
\newblock The medical segmentation decathlon.
\newblock {\em arXiv preprint arXiv:2106.05735}, 2021.

\bibitem{acdc}
Olivier Bernard, Alain Lalande, Clement Zotti, Frederick Cervenansky, Xin Yang,
  Pheng-Ann Heng, Irem Cetin, Karim Lekadir, Oscar Camara, Miguel
  Angel~Gonzalez Ballester, et~al.
\newblock Deep learning techniques for automatic mri cardiac multi-structures
  segmentation and diagnosis: is the problem solved?
\newblock {\em IEEE transactions on medical imaging}, 37(11):2514--2525, 2018.

\bibitem{pipo}
Xi~Fang and Pingkun Yan.
\newblock Multi-organ segmentation over partially labeled datasets with
  multi-scale feature abstraction.
\newblock {\em IEEE Transactions on Medical Imaging}, 39(11):3619--3629, 2020.

\bibitem{focusnet}
Yunhe Gao, Rui Huang, Ming Chen, Zhe Wang, Jincheng Deng, Yuanyuan Chen, Yiwei
  Yang, Jie Zhang, Chanjuan Tao, and Hongsheng Li.
\newblock Focusnet: imbalanced large and small organ segmentation with an
  end-to-end deep neural network for head and neck ct images.
\newblock In {\em International Conference on Medical Image Computing and
  Computer-Assisted Intervention}, pages 829--838. Springer, 2019.

\bibitem{densevnet}
Eli Gibson, Francesco Giganti, Yipeng Hu, Ester Bonmati, Steve Bandula,
  Kurinchi Gurusamy, Brian Davidson, Stephen~P Pereira, Matthew~J Clarkson, and
  Dean~C Barratt.
\newblock Automatic multi-organ segmentation on abdominal ct with dense
  v-networks.
\newblock {\em TMI}, 37(8):1822--1834, 2018.

\bibitem{swinunetr}
Ali Hatamizadeh, Vishwesh Nath, Yucheng Tang, Dong Yang, Holger Roth, and
  Daguang Xu.
\newblock Swin unetr: Swin transformers for semantic segmentation of brain
  tumors in mri images.
\newblock {\em arXiv preprint arXiv:2201.01266}, 2022.

\bibitem{unter}
Ali Hatamizadeh, Yucheng Tang, Vishwesh Nath, Dong Yang, Andriy Myronenko,
  Bennett Landman, Holger~R Roth, and Daguang Xu.
\newblock Unetr: Transformers for 3d medical image segmentation.
\newblock In {\em Proceedings of the IEEE/CVF Winter Conference on Applications
  of Computer Vision}, pages 574--584, 2022.

\bibitem{kits}
Nicholas Heller, Niranjan Sathianathen, Arveen Kalapara, Edward Walczak, Keenan
  Moore, Heather Kaluzniak, Joel Rosenberg, Paul Blake, Zachary Rengel, Makinna
  Oestreich, et~al.
\newblock The kits19 challenge data: 300 kidney tumor cases with clinical
  context, ct semantic segmentations, and surgical outcomes.
\newblock {\em arXiv preprint arXiv:1904.00445}, 2019.

\bibitem{nnunet}
Fabian Isensee, Paul~F Jaeger, Simon~AA Kohl, Jens Petersen, and Klaus~H
  Maier-Hein.
\newblock nnu-net: a self-configuring method for deep learning-based biomedical
  image segmentation.
\newblock {\em Nature methods}, 18(2):203--211, 2021.

\bibitem{uxnet}
Yuanfeng Ji, Ruimao Zhang, Zhen Li, Jiamin Ren, Shaoting Zhang, and Ping Luo.
\newblock Uxnet: searching multi-level feature aggregation for 3d medical image
  segmentation.
\newblock In {\em International Conference on Medical Image Computing and
  Computer-Assisted Intervention}, pages 346--356. Springer, 2020.

\bibitem{chaos}
A~Emre Kavur, N~Sinem Gezer, Mustafa Bar{\i}{\c{s}}, Sinem Aslan, Pierre-Henri
  Conze, Vladimir Groza, Duc~Duy Pham, Soumick Chatterjee, Philipp Ernst,
  Sava{\c{s}} {\"O}zkan, et~al.
\newblock Chaos challenge-combined (ct-mr) healthy abdominal organ
  segmentation.
\newblock {\em Medical Image Analysis}, 69:101950, 2021.

\bibitem{btcv}
Bennett Landman, Z~Xu, J~Igelsias, M~Styner, T~Langerak, and A~Klein.
\newblock Multi-atlas labeling beyond the cranial vault-workshop and challenge.
\newblock 2017.

\bibitem{hdenseunet}
Xiaomeng Li, Hao Chen, Xiaojuan Qi, Qi~Dou, Chi-Wing Fu, and Pheng-Ann Heng.
\newblock H-denseunet: hybrid densely connected unet for liver and tumor
  segmentation from ct volumes.
\newblock {\em IEEE transactions on medical imaging}, 37(12):2663--2674, 2018.

\bibitem{word}
Xiangde Luo, Wenjun Liao, Jianghong Xiao, Tao Song, Xiaofan Zhang, Kang Li,
  Guotai Wang, and Shaoting Zhang.
\newblock Word: Revisiting organs segmentation in the whole abdominal region.
\newblock {\em arXiv preprint arXiv:2111.02403}, 2021.

\bibitem{adb1k}
Jun Ma, Yao Zhang, Song Gu, Cheng Zhu, Cheng Ge, Yichi Zhang, Xingle An,
  Congcong Wang, Qiyuan Wang, Xin Liu, et~al.
\newblock Abdomenct-1k: Is abdominal organ segmentation a solved problem.
\newblock {\em TPAMI}, 2021.

\bibitem{vnet}
Fausto Milletari, Nassir Navab, and Seyed-Ahmad Ahmadi.
\newblock V-net: Fully convolutional neural networks for volumetric medical
  image segmentation.
\newblock In {\em 2016 fourth international conference on 3D vision (3DV)},
  pages 565--571. IEEE, 2016.

\bibitem{nsd}
Stanislav Nikolov, Sam Blackwell, Alexei Zverovitch, Ruheena Mendes, Michelle
  Livne, Jeffrey De~Fauw, Yojan Patel, Clemens Meyer, Harry Askham, Bernardino
  Romera-Paredes, et~al.
\newblock Deep learning to achieve clinically applicable segmentation of head
  and neck anatomy for radiotherapy.
\newblock {\em arXiv preprint arXiv:1809.04430}, 2018.

\bibitem{pytorch}
Adam Paszke, Sam Gross, Francisco Massa, Adam Lerer, James Bradbury, Gregory
  Chanan, Trevor Killeen, Zeming Lin, Natalia Gimelshein, Luca Antiga, Alban
  Desmaison, Andreas Kopf, Edward Yang, Zachary DeVito, Martin Raison, Alykhan
  Tejani, Sasank Chilamkurthy, Benoit Steiner, Lu~Fang, Junjie Bai, and Soumith
  Chintala.
\newblock Pytorch: An imperative style, high-performance deep learning library.
\newblock In {\em Advances in Neural Information Processing Systems 32}, pages
  8024--8035. Curran Associates, Inc., 2019.

\bibitem{unet}
Olaf Ronneberger, Philipp Fischer, and Thomas Brox.
\newblock U-net: Convolutional networks for biomedical image segmentation.
\newblock In {\em International Conference on Medical image computing and
  computer-assisted intervention}, pages 234--241. Springer, 2015.

\bibitem{covidseg}
Holger Roth, Ziyue Xu, Carlos~Tor Diez, Ramon~Sanchez Jacob, Jonathan Zember,
  Jose Molto, Wenqi Li, Sheng Xu, Baris Turkbey, Evrim Turkbey, et~al.
\newblock Rapid artificial intelligence solutions in a pandemic-the covid-19-20
  lung ct lesion segmentation challenge.
\newblock 2021.

\bibitem{cascade_unet}
Holger~R Roth, Hirohisa Oda, Xiangrong Zhou, Natsuki Shimizu, Ying Yang,
  Yuichiro Hayashi, Masahiro Oda, Michitaka Fujiwara, Kazunari Misawa, and
  Kensaku Mori.
\newblock An application of cascaded 3d fully convolutional networks for
  medical image segmentation.
\newblock {\em Computerized Medical Imaging and Graphics}, 66:90--99, 2018.

\bibitem{oan}
Yan Wang, Yuyin Zhou, Wei Shen, Seyoun Park, Elliot~K Fishman, and Alan~L
  Yuille.
\newblock Abdominal multi-organ segmentation with organ-attention networks and
  statistical fusion.
\newblock {\em Medical image analysis}, 55:88--102, 2019.

\bibitem{step}
Mark Weber, Jun Xie, Maxwell Collins, Yukun Zhu, Paul Voigtlaender, Hartwig
  Adam, Bradley Green, Andreas Geiger, Bastian Leibe, Daniel Cremers, et~al.
\newblock Step: Segmenting and tracking every pixel.
\newblock {\em arXiv preprint arXiv:2102.11859}, 2021.

\bibitem{cotr}
Yutong Xie, Jianpeng Zhang, Chunhua Shen, and Yong Xia.
\newblock Cotr: Efficiently bridging cnn and transformer for 3d medical image
  segmentation.
\newblock In {\em International conference on medical image computing and
  computer-assisted intervention}, pages 171--180. Springer, 2021.

\bibitem{c2fnas}
Qihang Yu, Dong Yang, Holger Roth, Yutong Bai, Yixiao Zhang, Alan~L Yuille, and
  Daguang Xu.
\newblock C2fnas: Coarse-to-fine neural architecture search for 3d medical
  image segmentation.
\newblock In {\em Proceedings of the IEEE/CVF Conference on Computer Vision and
  Pattern Recognition}, pages 4126--4135, 2020.

\bibitem{itk}
Paul~A Yushkevich, Joseph Piven, Heather~Cody Hazlett, Rachel~Gimpel Smith,
  Sean Ho, James~C Gee, and Guido Gerig.
\newblock User-guided 3d active contour segmentation of anatomical structures:
  significantly improved efficiency and reliability.
\newblock {\em Neuroimage}, 31(3):1116--1128, 2006.

\bibitem{dodnet}
Jianpeng Zhang, Yutong Xie, Yong Xia, and Chunhua Shen.
\newblock Dodnet: Learning to segment multi-organ and tumors from multiple
  partially labeled datasets.
\newblock In {\em Proceedings of the IEEE/CVF Conference on Computer Vision and
  Pattern Recognition}, pages 1195--1204, 2021.

\bibitem{nnformer}
Hong-Yu Zhou, Jiansen Guo, Yinghao Zhang, Lequan Yu, Liansheng Wang, and Yizhou
  Yu.
\newblock nnformer: Interleaved transformer for volumetric segmentation.
\newblock {\em arXiv preprint arXiv:2109.03201}, 2021.

\bibitem{psnn}
Yuyin Zhou, Zhe Li, Song Bai, Chong Wang, Xinlei Chen, Mei Han, Elliot Fishman,
  and Alan~L Yuille.
\newblock Prior-aware neural network for partially-supervised multi-organ
  segmentation.
\newblock In {\em Proceedings of the IEEE/CVF International Conference on
  Computer Vision}, pages 10672--10681, 2019.

\bibitem{coarse2fine}
Zhuotun Zhu, Yingda Xia, Wei Shen, Elliot Fishman, and Alan Yuille.
\newblock A 3d coarse-to-fine framework for volumetric medical image
  segmentation.
\newblock In {\em 2018 International conference on 3D vision (3DV)}, pages
  682--690. IEEE, 2018.

\end{thebibliography}
\newpage

\clearpage
\appendix
\begin{center}
	\LARGE \bf {Appendix}
\end{center}
\etocdepthtag.toc{mtappendix}
\etocsettagdepth{mtchapter}{none}
\etocsettagdepth{mtappendix}{subsection}
\tableofcontents
\section{Abdominal Organ Segmentation Datasets}
\label{app:datasets}
In this part, we provide detailed descriptions of previous abdominal organ segmentation datasets. We first introduce the datasets covering single organs in Sec.~\ref{single-organ}. The introductions of multi-organs Datasets will be developed in Sec.~\ref{multi-organ}. The statistics of the specific datasets are summarized in \cref{tab:spacing}.

\subsection{Single-organ Datasets} \label{single-organ}
\paragraph{MSD-Liver dataset~\cite{msd}} consists of 131 training and 70 testing CT cases with liver and liver tumor annotations. These scans were collected at 7 medical centers with patients suffering from various primary cancers.

\paragraph{MSD-Spleen dataset~\cite{msd}} provides 61 cases with spleen annotations, which are provided by the Memorial Sloan Kettering Cancer Center (New York, USA) with patients undergoing chemotherapy treatment for liver metastases. The annotations are first generated using a level-set-based method semi-automatically, and finally revised by an expert abdominal radiologist.

\paragraph{MSD-Prostate dataset~\cite{msd}} maintains 48 prostate multiparametric MRI (mpMRI) with the annotations covering the prostate peripheral zone and the transition zone. The data was acquired at Radboud University Medical Center, Nijmegen Medical Centre, Nijmegen, the Netherlands.

\paragraph{MSD-Pancreas dataset ~\cite{msd}} contains 420 patients suffering from pancreatic masses in Memorial Sloan Kettering Cancer Center (New York, USA).
The pancreatic parenchyma and pancreatic mass (i.e., cyst or tumor) annotations are provided, which are manually annotated by an radiologist.

\paragraph{KiTS dataset~\cite{kits}} includes 300 cases with kidney and kidney tumor annotations, which are acquired at the University of Minnesota Medical Center (Minnesota, USA). The patients in this dataset are suffered from kidney cancer. The kidney and tumor annotations were segmented by junior medical students under the supervision of a clinical chair.

\subsection{Multi-organ Datasets} \label{multi-organ}

\paragraph{BTCV dataset~\cite{btcv}} consists of 50 abdominal CT scans acquired at the Vanderbilt University Medical Center from metastatic liver cancer patients or post-operative ventral hernia patients. 
This benchmark aims to segment 13 organs, including the spleen, right kidney, left kidney, gallbladder, esophagus, liver, stomach, aorta, inferior vena cava, portal and splenic vein, pancreas, right adrenal gland, and left adrenal gland.
The organs were manually labelled by two experienced undergraduate students, and verified by a radiologist.

\paragraph{Chaos dataset~\cite{chaos}} consists of 20 CT scans with liver annotations and 20 MRI cases with four organ annotations (i.e., liver, spleen, left kidney, right kidney), which are collected by Dokuz Eylul University (DEU) hospital (˙Izmir, Turkey). The samples in this dataset are acquired from a healthy population.

\paragraph{DenseVNet dataset~\cite{densevnet}} comprise 90 abdominal CT images and the corresponding segmentation masks of 8 organs.
The cases are collected from the Cancer Imaging Archive (TCIA) Pancreas-CT dataset with pancreas segmentation, and the Beyond the Cranial Vault (BTCV) challenge with the segmentation of all organs except the duodenum. 
An imaging research fellow manually labeled the unsegmented organs under the supervision of a board-certified radiologist.

\paragraph{AbdomentCT-1K dataset~\cite{adb1k}} is a dataset with 1132 cases covering the liver, kidney, and pancreas annotation, which consists of 1112 3D CT scans from five existing datasets, including MSD-Liver (201 cases), KiTS (300 cases), MSD Spleen (61 cases) and Pancreas (420 cases), NIH Pancreas (80 cases), and a new dataset from Nanjing University (50 cases). Specifically, the overall 50 CT scans in the Nanjing University dataset are from 20 patients with pancreas cancer, 20 with colon cancer, and 10 with liver cancer. Annotations from the existing datasets are used if available. Besides, the absent organs will be further annotated in these datasets.

\paragraph{Word dataset~\cite{word}} consists of 150 cases with 16 types of organ annotations. The scans are collected from patients who had the prostatic cancer, cervical cancer, or rectal cancer. The annotations were manually labeled from scratch.

Slice spacing and number of slices refer to the 3D image axial plane resolution and spacing

\begin{table}[!tp]\centering
\scriptsize
\begin{tabular}{lrrrrrrr}\hline
&\multicolumn{4}{c}{Intensivty Property} &\multicolumn{2}{c}{Spatial Property} \\\cline{2-7}
&median &mean &5\% &99.50\% &slice spacing (mix/max/median) &slice num range (mix/max/median) \\
MSD-Liver &101 &99.39 &-17 &201 &[0.70 / 5.0 / 1.0] &[74 / 987 / 432.0] \\
MSD-Spleen &105 &99.29 &-41 &176 &[1.5 / 8.0 / 5.0] &[31 / 168 / 90.0] \\
MSD-Prostate & 641 & 854.69 &0 &2186 &[3.0 / 4.0 / 3.6] &[11 / 24 / 20] \\
MSD-Pancreas &84 &77 &-96 &215 &[0.70 / 7.5 / 2.5] &[37 / 751 / 93.0] \\

Kits &100 &100 &-79 &303 &[0.43 / 1.04 / 0.78] &[512 / 796 / 512] \\
BTCV &96 &83 &-958 &326 &[2.5 / 5.0 / 3.0] &[85 / 198 / 127] \\
Chaos &325 &361.38 &40 &1081 &[5.5 / 9.0 / 9.0] & [26 / 50 / 30] \\
DenseVNet&32&-46&-1003 &443 &[1.25 / 5.0 / 2.5] &[37 / 174 / 93] \\ \hline
\ours &- &- &- &- &[0.82 / 6.0 / 5.0] &[40 / 535 / 115.0] \\
\amosct &57 &50 &-991 &362 &[1.25 / 5.0 / 5.0] &[67 / 369 / 115.0] \\
\amosmri &768 &25383 &32 &164273 &[0.82 / 6.0 / 2.0] &[40 / 535 / 115.0] \\
\hline
\end{tabular}
\vspace{0.15cm}
\caption{Intensity and Spatial statistics of the conventional abdominal organ segmentation datasets. \revise{The slice spacing and number of slices denote the axial plane spacing and resolution of the images}}
\label{tab:spacing}
\end{table}

\section{\ours: Additional Details}
\label{app:amos}

\subsection{Data Acquisition}
All data in \ours are collected from eight scanners with different brands.
Acquisition details are different for each institution since they follow different clinical protocols in the clinical scenario. For example, 50 CT scans collected from the same scanner are obtained via the criteria of 120 kVP tube, 500 mm data collection diameter, 500-800 ms exposure time,  and 50-400 mA Xray tube current. Images were reconstructed at the 2.5-5 mm section thickness with a standard FC08 convolutional kernel and a 400-500 mm reconstruction diameter.
All data contributions to this study have been reviewed and approved by the Research Ethics Committee of Longgang District People's Hospital (reference number: 2021077) and the Research Ethics Committee of Longgang District Central Hospital (reference number: 2021ECJ012). The approved documents can be found in \url{https://drive.google.com/drive/folders/1UNHjEgau85rit-DBAKg9kGv6REkiHU6Z?usp=sharing}.

\begin{figure*}[t!]
	\centering
	\includegraphics[width=\linewidth]{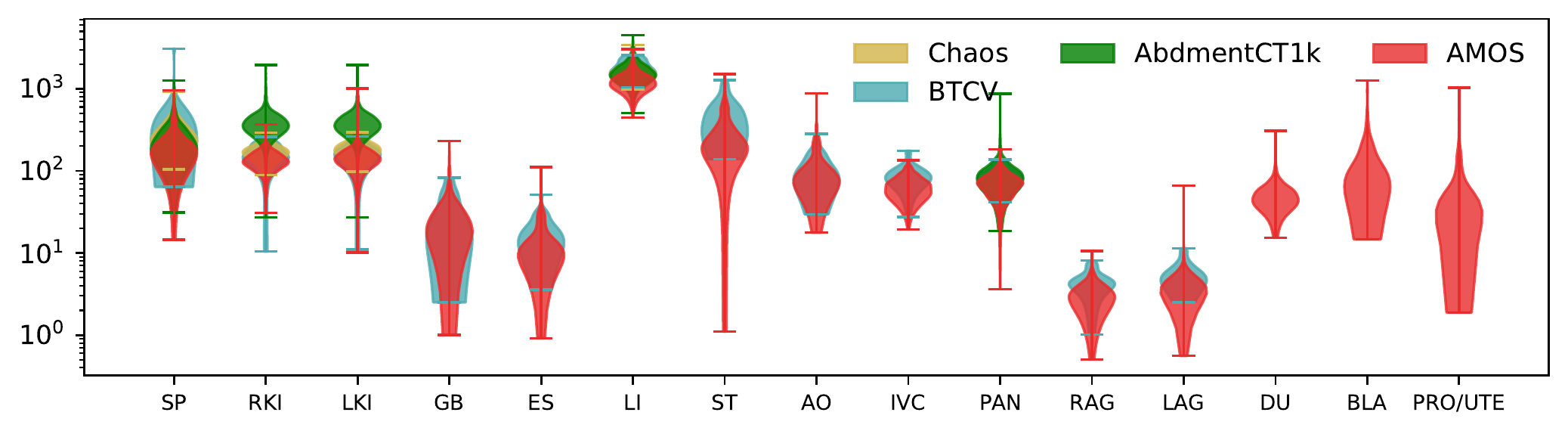}
	\caption{Organ volume distribution of BTCV, Chaos, AbdomentCT-1K and \ours datasets.}
	\label{fig:amos:comparsion:volume}
\end{figure*}

\subsection{Data Annotation}
\label{app:amos:annotate}
\ours adopts a semi-automatic annotation workflow as shown in \cref{fig:amos:annotation_workflow} in the main paper.
The coarse masks generated by a pre-trained segmentation model are first refined by five junior radiologists with 5 years of experience in clinical scanning, and further supervised by three board-certified senior radiologists with 10 years of experiences.
The segmentation labeling was performed slice-by-slice in the sagittal plane. Besides, the volumetric consistency was enforced by correcting segmentation in the axial and coronal planes in the ITK-SNAP~\cite{itk} toolbox. Each scan is annotated by one single annotator without multiple annotations for aggregation.
\revise{For the consistency of the annotation review, specifically, each senior physician will first individually review and record their comments, including a description of the problem and the corresponding image location, and then the comments will be aggregated and discussed to reach a final consensus opinion.}

\subsection{Data Distribution, Hosting, and Maintenance}
\label{app:amos:dist}
All data is distributed under the CC BY-NC-SA (Attribution-NonCommercial-ShareAlike) license. Data is hosted on the AWS open data platform and maintained by the authors.
Instructions for downloading and using the dataset can be found in the page \url{https://amos22.grand-challenge.org/}.
Further, we will establish a github repository to solicit possible annotation errors from data users.

\begin{table}[!h]\centering
\scriptsize
\begin{tabular}{lrrrrrrr}\hline
&\multicolumn{4}{c}{Intensivty Property} &\multicolumn{2}{c}{Spatial Property} \\\cline{2-7}
&median &mean &5\% &99.50\% &slice spacing (mix/max/median) &slice num range (mix/max/median) \\
\amosct-A & 69 & 63.76 & -967 & 403 & [5.0/ 5.0/ 5.0] &[68/ 140/ 98] \\
\amosct-B & 45 & 25 & -989 & 158 & [5.0/ 5.0/ 5.0] &[68/ 112/ 93] \\
\amosct-C & 63 & 53.68 & -996 & 401 & [1.25 / 5.0 / 2.0] &[76/ 353/ 213] \\
\rowcolor{gray!25}
\amosct-D & 61.0 & 54 & -994 & 337 & [1.25 / 5.0 / 2.0] &[78 / 321 / 203] \\
\rowcolor{gray!25}
\amosct-E & 59 & 56 & -979 & 353 & [1.25 / 5.0 / 2.0] &[67 / 369 / 224] \\  \hline
\rowcolor{gray!25}
\amosmri-F & 57422 & 58761 & 59 & 170721 & [0.86/ 6.0 / 1.40] &[60 / 535 / 320] \\
\amosmri-G & 2560 &344  & 16 & 66331 & [0.82 / 3.0 / 3.0] &[64 / 512 / 72] \\
\rowcolor{gray!25}
\amosmri-H & 845 &22937  & 47 & 1451508 & [0.82 / 2.0 / 0.82] &[100 / 512 / 512] \\ \hline
\end{tabular}
\vspace{0.15cm}
\caption{Intensity and spatial statistics of data generated from different scanners. Unseen test data are marked as gray. \revise{The slice spacing and number of slices denote the axial plane spacing and resolution of the images}}
\vspace{-0.15cm}
\label{tab:spacing:scanners}
\end{table}

\subsection{Data Statistics}
\label{app:amos:stat}
Data were collected from patients with abdominal tumors (majority) or other abnormalities. Moreover, among the 500 CT scans collected, the number of males and females are 314 and 186, respectively. For the age distribution, the patients' minimum, maximum, median, and mean ages are 14, 94, 54, and 53.64 years old, respectively. For the 100 MRI scans, the number of males and females are 55 and 45, and the patients' minimum, maximum, median, and mean ages are 22, 85, 50, and 48.71 years old, respectively. The ratio between the number of patients diagnosed with tumors and the number of patients with other abnormalities is 3:2. We manually set the distribution of these factors consistent between the training/validate/test splits.
We also analyze the intensity and spatial property of the data collected from different scanners and summarize them in \cref{tab:spacing:scanners}.

\section{Experiment}
In this section, we present the experimental details covering the model architectures, training schedules and so on. We follow the nnUNet package \cite{nnunet} to conduct the model training and evaluation in Pytorch \cite{pytorch}.
The code used to produce the results in our paper will be available at \url{https://github.com/JiYuanFeng/AMOS2022}.

\setlength{\tabcolsep}{1.5pt}
\renewcommand{\arraystretch}{1.1}
\begin{table}[!t]
	\scriptsize
	\centering
	\subtable
	{ 
	\begin{tabular}{l|c|ccccccccccccccc} \hline
		\multirow{2}{*}{Model} &\multirow{2}{*}{mDice $\uparrow$} &\multicolumn{15}{c}{Categorical Dice $\uparrow$} \\ \cline{3-17}
		&  &SPL &RKI &LKI &GBL &ESO &LIV &STO &AOR &IVC &PAN &RAG &LAG &DUO &BLA &PRO/UTE \\ \hline
        UNet \cite{unet}& 88.87& 96.31& 95.29& 96.28& 81.53& 85.72& 97.05& 90.77& 95.37& 91.53& 87.39& 79.83& 81.12& 82.56& 88.42& 83.81 \\
        VNet \cite{vnet}& 81.96& 94.21& 91.86& 92.65& 70.25& 79.04& 94.65& 84.79& 92.96& 87.4& 80.5& 72.62& 73.19& 71.69& 77.02& 66.62 \\
        CoTr \cite{cotr}& 77.13& 91.09& 87.18& 86.36& 60.47& 80.9& 91.61& 80.09& 93.66& 87.72& 76.32& 73.68& 71.74& 67.98& 67.38& 40.84 \\
        nnFormer \cite{nnformer}& 85.63& 95.91& 93.51& 94.8& 78.47& 81.09& 95.89& 89.4& 94.16& 88.25& 85.0& 75.04& 75.92& 78.45& 83.91& 74.58 \\
        UNetr \cite{unter}& 78.33& 92.68& 88.46& 90.57& 66.5& 73.31& 94.11& 78.73& 91.37& 83.99& 74.49& 68.15& 65.28& 62.35& 77.44& 67.52 \\
        Swin-UNetr \cite{swinunetr}& 86.37& 95.49& 93.82& 94.47& 77.34& 83.05& 95.95& 88.94& 94.66& 89.58& 84.91& 77.2& 78.35& 78.59& 85.79& 77.39 \\
		\hline
	\end{tabular}
	}
	\subtable
	{
	\begin{tabular}{l|c|ccccccccccccccc} \hline
		\multirow{2}{*}{Model} &\multirow{2}{*}{mNSD $\uparrow$} &\multicolumn{15}{c}{Categorical NSD $\uparrow$} \\ \cline{3-17}
		&  &SPL &RKI &LKI &GBL &ESO &LIV &STO &AOR &IVC &PAN &RAD &LAD &DUO &BLA &PRO/UTE \\ \hline
		UNet \cite{unet}& 79.87& 89.62& 88.89& 89.86& 72.91& 78.41& 83.9& 76.29& 90.61& 79.62& 73.37& 83.32& 83.25& 69.35& 76.16& 62.43 \\
		VNet \cite{vnet}& 67.94& 83.39& 81.14& 83.59& 58.43& 65.56& 74.99& 63.49& 84.86& 69.62& 60.29& 74.23& 72.33& 52.63& 56.8& 37.7 \\
		CoTr \cite{cotr}& 64.15& 77.66& 75.27& 74.64& 48.97& 69.47& 71.91& 59.46& 86.34& 72.13& 56.82& 76.18& 70.68& 52.39& 54.52& 15.81 \\
		nnFormer \cite{nnformer}& 74.15& 87.73& 85.82& 87.41& 68.0& 69.85& 80.88& 72.3& 86.82& 71.73& 68.21& 77.95& 77.32& 61.09& 67.7& 49.39 \\
		UNetr \cite{unter}& 61.49& 77.58& 76.06& 77.36& 50.78& 58.66& 72.09& 51.45& 78.89& 60.91& 51.91& 69.74& 63.23& 41.16& 55.28& 37.3 \\
		Swin-UNetr \cite{swinunetr}& 75.32& 87.46& 85.7& 86.76& 67.34& 73.62& 80.81& 71.17& 88.85& 74.99& 68.36& 80.52& 79.65& 61.97& 69.38& 53.2 \\
		\hline
	\end{tabular}
	}
    \caption{The class-wise scores on the validation set of \amosct dataset.
    }
	\label{tab:app:baseline}
	\vspace{-0.5cm}
\end{table}

\subsection{Implementation Details}
\label{app:exp:imp}

\paragraph{Data Prepossessing}

Following \cite{nnunet}, for the CT data, we first clip the HU values of each scans to the [-991, 362] range and then normalize truncated voxels values by subtracting 50 and dividing by 141. As for the MRI data, we adopt Z-score data normalization.

\paragraph{Baselines}
We benchmark various state-of-the-art medical segmentation methods. Unless otherwise specified, we follow the default configurations in their released codebases.
The implementation of these methods can be found in: UNet\footnote{ \url{https://github.com/MIC-DKFZ/nnUNet/blob/master/nnunet/network_architecture}}, VNet\footnote{\url{https://github.com/Project-MONAI/MONAI/blob/dev/monai/networks/nets} \label{refnote}}, CoTr\footnote{\url{https://github.com/YtongXie/CoTr/blob/main/CoTr_package/CoTr}}, nnFormer\footnote{\url{https://github.com/282857341/nnFormer/blob/main/nnformer/}}, UNetr\textsuperscript{\ref{refnote}}, Swin-UNetr\textsuperscript{\ref{refnote}}.

\paragraph{Training Schedule}
\setlength{\tabcolsep}{5pt}
\renewcommand{\arraystretch}{1}
\begin{wraptable}{r}{5.5cm}
\vspace{-0.3cm}
	\scriptsize
    \centering
    \begin{tabular}{lcc}
    \hline Parameter & Prob & Param \\
    \hline Random Rotation & $0.2$ & [-0.52, 0.52] \\
    Random Scale & $0.2$ & [0.70, 1.40] \\
    Random Gaussian-Noise & $0.1$ & [0.00, 0.10] \\
    Random Gaussian-Blur & $0.2$ &  [0.50, 1.00] \\
    Random Brightness & $0.15$ & [0.75, 1.25] \\
    Random Contrast & $0.15$ & [0.75, 1.25] \\
    Simulate Low-Resolution & $0.25$ & [0.50, 1.00] \\
    Random Gamma & $0.3$ & [0.7, 1.5] \\ 
    Random Mirror & 1 & \\
    \hline
    \end{tabular}
    \caption{Parameters of the used data augmentations}
    \label{tab:app:agumentation}
\vspace{-0.5cm}
\end{wraptable}
In the training stage, we randomly crop sub-volume sizes to 64 $\times$ 160 $\times$ 160 and 48 $\times$ 160 $\times$ 224 for CT, and MRI scans as input, respectively.
All the experiments are conducted using 1 NVIDIA V100 GPU with a batch size of 2.
For data augmentation, we follow the configurations in \cite{nnunet}, including random rotation, scaling, flipping, Gaussian noising, Gaussian blurring, brightness and contrast adjusting, simulation of low resolution, and Gamma transformation. The detailed augmentation parameters are listed in \cref{tab:app:agumentation}.
We train each model for the same 1000 epochs for fair comparisons.
For network optimization, we configured the training objective as the combination of cross-entropy loss and dice loss. Besides, we adopt the SGD algorithm with a momentum of 0.99 and an initial learning rate of 0.01 as the optimizer.

In the testing stage, we employ the sliding window inference strategy where the window sizes equal the training patch size. Besides, data augmentation, like flipping, is also utilized in the testing process. To quantitatively evaluate the segmentation results, we calculate the Dice Similarity Coefficient (DSC) and Normalized Surface Dice (NSD) scores. A higher score indicates a better segmentation performance.

\subsection{Additional Results}
\label{app:exp:res}

\paragraph{Class-wise results}
We provide the detailed class-wise scores of the benchmarked methods on the validation set in \cref{tab:app:baseline}. The corresponding abbreviations are presented as follows:
spleen (SPL), right kidney (RKI), left kidney (LKI), gallbladder (gbl), esophagus (ESO), liver (LIV), stomach (STO), aorta (AOR), inferior vena cava (IVC), pancreas (PAN), right adrenal gland (RAG), left adrenal gland (LAG), duodenum (DUO), bladder (BLA), prostate/uterus (PRO/UTE).

\paragraph{Transfer learning}
We perform the task-transfer by fine-tuning the pre-trained models on the ten medical segmentation datasets, including six related datasets containing organ annotations in \ours, and four unrelated ones.
Information about the datasets are summarized in \cref{tab:app:transfer:dataset}.
We adopt the standard fine-tuning protocols, where we initial the network with the parameters of the pre-trained representation from \ours. We apply the same training and testing schedule as introduced in \cref{app:exp:imp}.

\setlength{\tabcolsep}{2pt}
\renewcommand{\arraystretch}{1}
\begin{table}[!t]\centering
\scriptsize
\begin{tabular}{lcclccl|cl}\hline
Dataset &Modality &Classes &Description &Available &Train &Val & DS Str & Patch \\ \hline
MSD-Liver &CT &2 &Liver and tumour &131 &104 & 27 & [5, 5, 5] & [128, 128, 128] \\
MSD-Spleen &CT &1 &Spleen &41 &32 &9 &[4, 5, 5] &[ 64, 192, 160] \\
MSD-Pancreas &CT &2 &Pancreas and tumour &281 &224  &57 &[3, 5, 5] &[40, 224, 224] \\
MSD-Prostate &MRI &2 &Prostate central gland and Peripheral zone &32 &25 &7 &[2, 6, 6] &[20, 320, 256] \\ 
MSD-Kits &CT &2 &Kidney and tumour &210 & 168 &42 &[5, 5, 5] &[128, 128, 128]\\ \hline
MSD-Cardiac &MRI &1 &Left Atrium &20 &16 &4 &[4, 5, 5] &[ 80, 192, 160] \\
MSD-HepaticVessel &CT &2 &Hepatic vessels and tumour &303 &242 &61 &[4, 5, 5] &[64, 192, 192] \\
ACDC &MRI &4 & RV, MLV, LVC &200 &160 &40 &[2, 5, 5]&[20, 256, 224]  \\
Covid-19 &CT &1 &Covid Lesion & 199 &159 &40 & [2, 6, 6] & [28, 256, 256]  \\ 
SegTHOR  &CT &1 &Esophagus, Heart, Trachea, Aorta  & 40 &32 &8 & [4, 5, 5] &[64, 192, 160] \\ 
\hline
\end{tabular}
\vspace{0.2cm}
\caption{Characteristics of the datasets used in transfer learning. Based on the available data, We divide the training and validation set according to the ratio of 8:2. Besides, we report the downsample stride (abbreviated as DS Str) of the used UNet architecture configuration, as well as the input patch size of each task.}
\label{tab:app:transfer:dataset}.
\end{table}





\end{document}